# Synthesis of Length-Tunable DNA Carriers for Nanopore Sensing

Zachary Roelen, Vincent Tabard-Cossa*

Department of Physics, University of Ottawa, Ottawa, Ontario, Canada

*Corresponding author: tcossa@uottawa.ca

**ABSTRACT**

Molecular carriers represent an increasingly common strategy in the field of nanopore sensing to use secondary molecules to selectively report on the presence of target analytes in solution, allowing for sensitive assays of otherwise hard-to-detect molecules such as small, weakly-charged proteins. However, existing carrier designs can often introduce drawbacks to nanopore experiments including higher levels of cost/complexity and carrier-pore interactions that lead to ambiguous signals and elevated clogging rates. In this work, we present a simple method of carrier production based on sticky-ended DNA molecules that emphasizes ease-of-synthesis and compatibility with nanopore sensing and analysis. In particular, our method incorporates the ability to flexibly control the length of the DNA carriers produced, enhancing the multiplexing potential of this carrier system through the separable nanopore signals they could generate for distinct targets. A proof-of-concept nanopore experiment is also presented, involving carriers produced by our method with multiple lengths and attached to DNA nanostructure targets, in order to validate the capabilities of the system. As the breadth of applications for nanopore sensors continues to expand, the availability of tools such as those presented here to help translate the outcomes of these applications into robust nanopore signals will be of major importance.

**INTRODUCTION**

Molecular carriers are a sub-category of an increasingly popular trend in nanopore sensing to use designed, structured molecules to characterize a sample of biological interest [3] or to probe the fundamentals of polymer physics [4]. Under the molecular carrier scheme, instead of sensing a target analyte directly with a nanopore, a proxy molecule ("carrier") that interacts with this target (or with intermediate reporters of the target) is sensed. The key requirement of this approach is for the nanopore signals generated by the translocating carriers to be recognizably different depending on whether this interaction with the target has occurred or not. For instance, a target could bind to a specific subregion of a carrier, resulting in this subregion blocking additional ionic current when passing through the pore (as compared to without the target attached) [5]. In this way, the presence of the target analyte can be observed indirectly through its effect on the carrier.

Using molecular carriers in place of direct sensing can provide a number of benefits, depending on the application. For one, it can increase the sensitivity of a nanopore experiment for a particular target, such as in cases where the target molecule on its own is not easily captured (e.g. weakly charged) or detected (e.g. small size, fast translocations) by the pore. It also offers a means of imparting target specificity to an experiment, without the need for complex functionalization of the nanopore itself. Many potential targets (e.g. globular proteins) can have very different functions (or diagnostic values in an assay) from each other, all while producing fairly similar (fast, structureless) signals when passing through a pore [6]. If, however, only one of the targets interacts in a unique way with a molecular carrier designed for this purpose, then its specific presence in a mix of otherwise indistinguishable molecules can be established.

Finally, molecular carriers can allow for nanopore sensors to probe a wide variety of target types without heavily modifying a base sensing protocol. An example of such a system could involve a collection of target molecules with varying values (in magnitude or in sign) of net electric charge. This again may be representative of globular proteins, for example, whose surface charges are pH-dependent and determined by the sequence of amino acids that compose them [7]. However, if every target was combined with a (e.g. highly-negatively charged) molecular carrier whose individual contribution dominated the net electrophoretic mobility of the target-carrier pair, then similar experimental conditions such as transmembrane potential could be used to detect each species. Single molecular carriers (or a mix of carriers) can also be engineered to have interaction probes for multiple targets, allowing a single nanopore experiment to report on each of these targets simultaneously ("multiplexing") [2,8].

A primary candidate for a carrier molecule is double-stranded DNA (dsDNA), which features high levels of regularly-spaced negative charge along its backbone and which has been extensively characterized inside the nanopore system [9]. In addition to being readily captured under an applied electric field, the rigid, double-helical structure of dsDNA results in it having a relatively open, extended conformation in solution [10], allowing for positional information (such as the attachment of a target molecule along its contour) to be read out in its translocation signal as it threads sequentially though the pore. To build a molecular carrier based on DNA, commercially-produced viral genomes are often used as the starting point. These have the advantages of being widely available and of possessing known sequences around which their target-selective functionalities can be designed. Here, we highlight two existing approaches to implementing DNA carriers based on viral genomes.

A first approach, following work by the Keyser group [1], starts with the circular, single-stranded genome of the M13 bacteriophage (7.25 knt in length, typically linearized first by enzymatic cleavage). Inspired by the staple strands of DNA origami [11], to this single-stranded genome are attached hundreds of shorter (~40 bp) complementary oligonucleotides in order to form a double-stranded construct. By modifying the sequence or chemical functionalization of particular oligos in the complement, the capacity to interact with specific targets can be integrated at precise locations along the contour of the completed carrier. Carriers built by this method have been used to detect and measure the concentration of multiple proteins [12], to distinguish two DNA strands differing by a single base [13], and to store digital data in the form of carrier structure [14–16]. An innovative way of signalling target specificity in this system is with the concept of "barcoding" where the particular target interaction being probed by a carrier is labelled by a nanopore-readable structure ("barcode") at a dedicated region in the sequence [8,16]. In this way, two targets that generate somewhat similar signals at the separate, interaction regions of their respective carriers can be reliably distinguished by the different readouts of their barcodes, while maintaining the aforementioned benefits of the carrier system as a whole.

A second, possibly simpler DNA carrier approach, as developed by the groups of Edel and Ivanov [2], is based on λ DNA which is a linear, double stranded genome (48.5 kbp) that features short, cohesive 12-nucleotide single-stranded extensions at its ends. It is to these cohesive ends that functionalized oligos imparting carrier capabilities to the final constructs can be attached, in analogy with the methods of Keyser & co. Carriers based on λ DNA have been used successfully to probe fluorescent DNA oligos [17], biotin-streptavidin binding [17], and protein-aptamer interactions [2,17]. Although the location of carrier activity within the DNA substrate is more restricted in this system (when compared to approaches involving fully single-stranded genomes and their 100+ oligo complements) each of the two cohesive ends can be effectively extended with linker oligos as needed and the simultaneous integration of up to three homo- and heterogeneous protein-aptamer pairs on a single λ-carrier has been demonstrated [2].

While molecular carriers produced from viral genomes can be versatile, there are some downsides associated with such sources of DNA. For one, an extended genome such as λ (48.5 kbp) represents a long length for a nanopore analyte (typically $\lesssim$ 10 kbp [9,18,19]). The longer a DNA molecule is, the more prone it is to entering the nanopore in a folded conformation [20] which could the cause the local, intra-event signals generated by substructures in the molecule (e.g. as from attached targets) to be read outside of their expected order [15]. Greater proportions of events going unrecognized in this way can necessitate longer experiment times (to generate sufficient statistics from the non-rejected events) and may even lead to experimental bias if correlations exist between the folding frequency of a carrier and its state of target attachment. Longer molecules also tend to irreversibly stall more often during a translocation attempt ("pore clogging"), potentially ending a single-pore experiment early before all sample conditions can be tested.

Additional issues can also arise from sourcing purified versions of these viral templates commercially. While widely available from popular vendors in several molecular configurations and order sizes, these purified genomes may become costly to purchase in the quantities needed to continuously run nanopore experiments, especially if losses are incurred in their subsequent processing into their final, modified forms. There is also variability that can exist (e.g. batch-to-batch, vendor-to-vendor) in the quality of a particular template sample, for instance in the fraction of normally circular genomes (e.g. M13) that can arrive randomly sheared (and therefore linearized) partway through their sequence [1]. This can translate into experimental variability if, after processing, the defective copies of the molecule present in the sample lead to altered nanopore signals.

The overall goal of this work is therefore to develop a protocol for the in-lab synthesis of DNA carriers that are well-suited for nanopore sensing. In particular, we focus on a design that follows the general structure of λ-DNA, namely, linear, double-stranded molecules that terminate in single-stranded overhangs ("sticky ends") onto which targets may be attached by sequence-specific DNA hybridization. Two main objectives are prioritized when developing the protocol. The first is to have any methods used be achievable with standard, countertop molecular biology lab equipment (e.g. thermocycler, gel electrophoresis system, microcentrifuge). This lowers the barrier for other researchers to make use of and iterate on these methods, without the need to procure access to highly-specialized, institution-shared equipment (e.g. HPLC system, ultracentrifuge). Note, however, that restricting ourselves to these common tools still allows for plenty of flexibility in the design and functionality of the molecular outputs. For instance, the powerful polymerase chain reaction (PCR) technique combines the thermocyler with a handful of widely-available consumables (including a thermostable DNA polymerase) to repeatably amplify DNA fragments with specific targeted sequences and can readily incorporate functional groups or labels into these fragments by simple substitutions in the starting materials (e.g. using modified dNTPs or primers) [21,22].

The second priority when designing our synthesis protocol is to allow the overall length of the sticky-ended carriers produced through these methods to be highly adjustable. As discussed above, a DNA fragment that is overly long can be problematic for nanopore sensing through its increased tendencies to adopt complex, difficult to analyze conformations and to clog pores. Conversely, there are also complications to sensing very short DNA fragments, arising from the finite bandwidth of the system used to measure the small nanopore currents (centred around a low-noise amplifier) – short fragments may translocate the membrane too quickly to be reliably detected or to have the details of their signals be well resolved by this system. Beyond the need for an intermediate fragment length between these two extremes however, there is additional utility in the context of molecular carriers to being able to finely adjust the product length within this range. This is because carrier length represents another variable by which carriers mapping to different assay targets could be separated by a nanopore, specifically through the dwell times or equivalent charge deficits of the translocation events they produce [18,23,24]. Moreover, many of the techniques for carrier identification highlighted above (such as barcoding or end-labelling) can be applied simultaneously with length selection on the same molecule, creating a multiplicative effect that supports high levels of multiplexing from a small number of variants in each mode.

In this work, we present a method for synthesizing sticky-ended DNA molecules (SE-DNA) that adheres to the objectives outlined above. The basic protocol is first outlined, with key aspects to its design highlighted that allow for the length of the product and the sequence of its sticky ends to be adjusted, as well as for compatibility with nanopore sensors to be maintained. Following on this, the capabilities of molecules produced by this method as molecular carriers are validated with a proof-of-concept nanopore experiment.

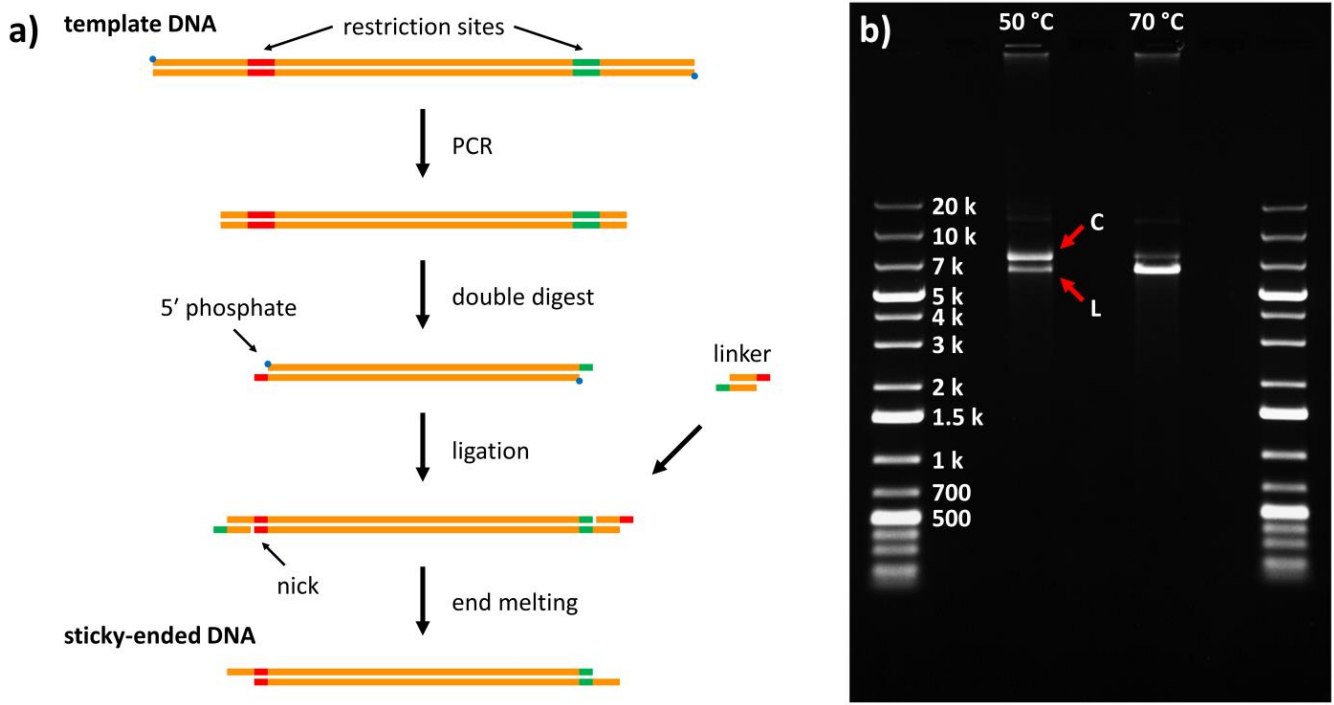

*Figure 1*: *a)* Overview of the basic SE-DNA synthesis protocol. The length of the final product is determined primarily by the distance between the two restriction sites (which can be adjusted as desired during the PCR step, see main text) while the number of sticky ends and their sequences are flexibly determined by the choice of linker molecule(s). *b)* Agarose gel (0.5%, 0.5× TAE, pre-stained with Gel Red, 70 V) of 6.2 kbp SE-DNA produced by this method, here using λ-DNA template, KpnI and SacI restriction enzymes, and a single linker. The second lane from the left ("50 °C") is the result of incubating the product for 1 hr at 50 °C (~1 nM DNA in 10 mM $MgCl_2$) and shows two main bands: the linear product ("L") migrating between the 5 kbp and 7 kbp DNA ladder bands, and a slower-migrating circular product ("C") resulting from the annealing of the two complementary ends of a linear molecule. The next lane over ("70 °C") is the result of taking an aliquot of the same "50 °C" mixture and heating it above the melting temperature of the 12-nt sticky ends (10 min at 70 °C) – here the circular band is greatly reduced in favour of the linear band as the annealed ends of the circles are reversibly melted.

**METHOD OUTLINE**

A schematic of the overall synthesis process of sticky-ended DNA molecules (SE-DNA) is presented in Figure 1a. It begins by selecting a DNA template with a known sequence (e.g. from a commercially-sourced bacteriophage genome) such that PCR primers can be designed to amplify a specific internal subregion of this sequence. After amplification, the product is digested with restriction endonucleases to create short (e.g. 4 nt) single-stranded overhangs on one or both ends of a central fragment. These ssDNA overhangs are targeted in a subsequent ligation step as the attachment sites of small double-stranded "linker" molecules.

The specific design of the linkers used in this step will define both the presence and the sequence of the sticky ends of the final product. In the example presented in Figure 1a, the PCR amplicon has been digested by two restriction enzymes to produce a different short ssDNA overhang on each end. The linker in this example then consists of two oligonucleotides, each of which features a complementary region to one of the two short overhang sequences (red and green segments in Figure 1a) followed by a region that is complementary to its analogue on the other linker strand, allowing for their hybridization into a single molecule. In this case, the sticky end sequences on the final product will be the same as those holding the two linker strands together, and will thus be cohesive with each other. However, by choosing to use one or two different linker molecules (two or four linker oligonucleotides) and by varying the lengths and specific sequences of their central regions, an essentially arbitrary ssDNA extension could be independently placed on one or both ends of the final product.

The final step in the process consists of melting off one of the linker strands on each end to expose the desired ssDNA extensions comprising the sticky ends. This relies on the fact that the linker molecules are not 5'-phosphorylated (as can be specified when ordering commercial chemically-synthesized DNA oligos) and so during the ligation step, only their 3' ends are covalently attached to the digested amplicon by the ligase. This leaves nicks placed near the ends of the products that lead to relatively unstable terminal stretches of annealed dsDNA (~10 bases long, exact length determined by the linker sequences, see Figure 1a) that can be selectively melted by heat or chemical denaturation, leaving the stable central region that composes the bulk of the molecule (~1000+ bases long) intact. Sections S1 and S2 in the Supporting Information contain the oligo sequences used in this work (primers and linkers), as well as detailed experimental protocols, respectively.

Referring back to the example of Figure 1a, we see that the length of the final sticky-ended product in this case is determined primarily by the distance between the recognition sites of the two restriction enzymes in its blunt precursor (PCR amplicon). An obvious method of modifying this length to meet the needs of an application could therefore be to target alternative recognition sites in the template sequence or even different restriction enzymes. There are, however, several downsides to such an approach. For one, the exact product lengths achievable this way are limited to only those that correspond to the locations of existing recognition sites (which may or may not be beneficially distributed in a given DNA template). Additionally, using different restriction enzymes on a template typically produces different cleaved ends, varying in overhang length (including 0 nt – blunt ends), polarity (5' or 3'), and/or sequence. This necessitates careful consideration that a targeted cleavage site is compatible with the construction of a particular product, or, at the very least, requires the re-design and purchase of additional linker molecules for each new cut type.

A more flexible approach to setting the product length might instead start with the selection of a single set of restriction enzymes whose recognition sites are rare or non-existent in the native sequence of the DNA template. New recognition sites for these enzymes can then be introduced at practically arbitrary locations within the template (thus defining the product length) by modification of the primers used in the PCR stage of synthesis – either by point mismatches between primer and template to create cleavage sites from closely-related sequences or by simple addition of the recognition sequence (typically ~6 bp) to the 5' end of the primer [25]. Section S3 in the Supporting Information presents an illustrative example involving the combination of KpnI and SacI restriction enzymes with a λ-DNA template, where the "forward" PCR primer is fixed near the location of a KpnI site and the "reverse" primer is placed somewhere within the region directly downstream (~17 kbp wide), which is devoid of SacI sites in native λ-DNA. Even when restricting the placement of the reverse primer to sequences within this region that differ by one base at most from a SacI site (and so could be converted to an active site with a single base substitution in the primer), a full range of product lengths from 0 – 17 kbp is achieved, with an average gap between adjacent lengths of ~350 bp (see Figure S3). Creating a wide range of different SE-DNA product sizes in this case then involves modifying a base synthesis protocol by only a single, low-cost PCR primer at a time.

Finally, we note the importance of removing any "extra" DNA fragments with single-stranded overhangs that appear at various stages of the synthesis protocol. These include the "caps" of the PCR amplicon that are cleaved off of the central fragment during the restriction digest, as well as any excess linker molecules that have not been incorporated onto this fragment during the ligation stage. If these spurious fragments are not removed, their presence in the reaction mixtures of subsequent stages could create interference by annealing with complementary regions on the main product and thus blocking interactions of these regions with their intended targets (see Figure 1a). So long as these extra fragments were relatively short (≲ 100 bp), their separation from the larger (1000+ kbp) main product was found to be easily achieved by incorporating a standard DNA purification step – e.g. by silica membrane-based spin column or alcohol precipitation – in between synthesis stages (see Section S4 in the Supporting Information). Crucially however, it was observed that any purification attempts that involved extracting DNA from agarose gels (as might be necessary for separating larger DNA fragments by size) created problems for the downstream nanopore sensing of these molecules. We found that samples that had interacted with agarose systematically led to increased pore clogging (both reversible and irreversible with voltage inversion or solution flushing) and additional, low-amplitude blockages in the current signal (as if from contaminants), when compared to control samples that lacked agarose purification (see Section S5 in the Supporting Information). It is therefore important to design any intermediate regions that will not make it into the final product (such as between a restriction site and the end of a PCR amplicon) to be relatively short and thus easily separable.

Figure 1b shows a gel image of a 6216-bp product created with the above protocol (using a λ-DNA template and KpnI/SacI restriction enzymes) that features two (5′, 12-nt long) cohesive ends, as in the example of Figure 1a. The first output lane (labelled "50 °C") shows the result of incubating a sample of this product at intermediate temperatures (50 °C), below the melting point of the sticky ends, for 1 hr (~1 nM DNA in 10 mM $MgCl_2$ + 10 mM Tris, pH 8). Two main populations are visible – a faster band migrating between the 5- and 7-kbp bands of the ladder that is attributed to the linear 6.2 kbp product, as well as a slightly slower-migrating band that is attributed to the two cohesive ends of the product annealing to form a circular molecule. A third, slower band is faintly visible above these two (between the 10- and 20-kbp bands of the ladder) and may represent rare dimers that were formed during the ligation step. In the adjacent output lane ("70 °C"), an aliquot of this same sample mixture was briefly heated at temperatures above the melting point of the 12-nt ends (10 min at 70 °C) and then immediately submerged in an ice bath. In this case, the "circular" band is greatly reduced in favour of the "linear" band, confirming that the sample contained at least a substantial proportion of products with two active sticky

ends – ends that were stably annealed to each other at intermediate temperatures and that could be reversibly dissociated (not locked-in during the ligation stage) with simple temperature cycling, as desired.

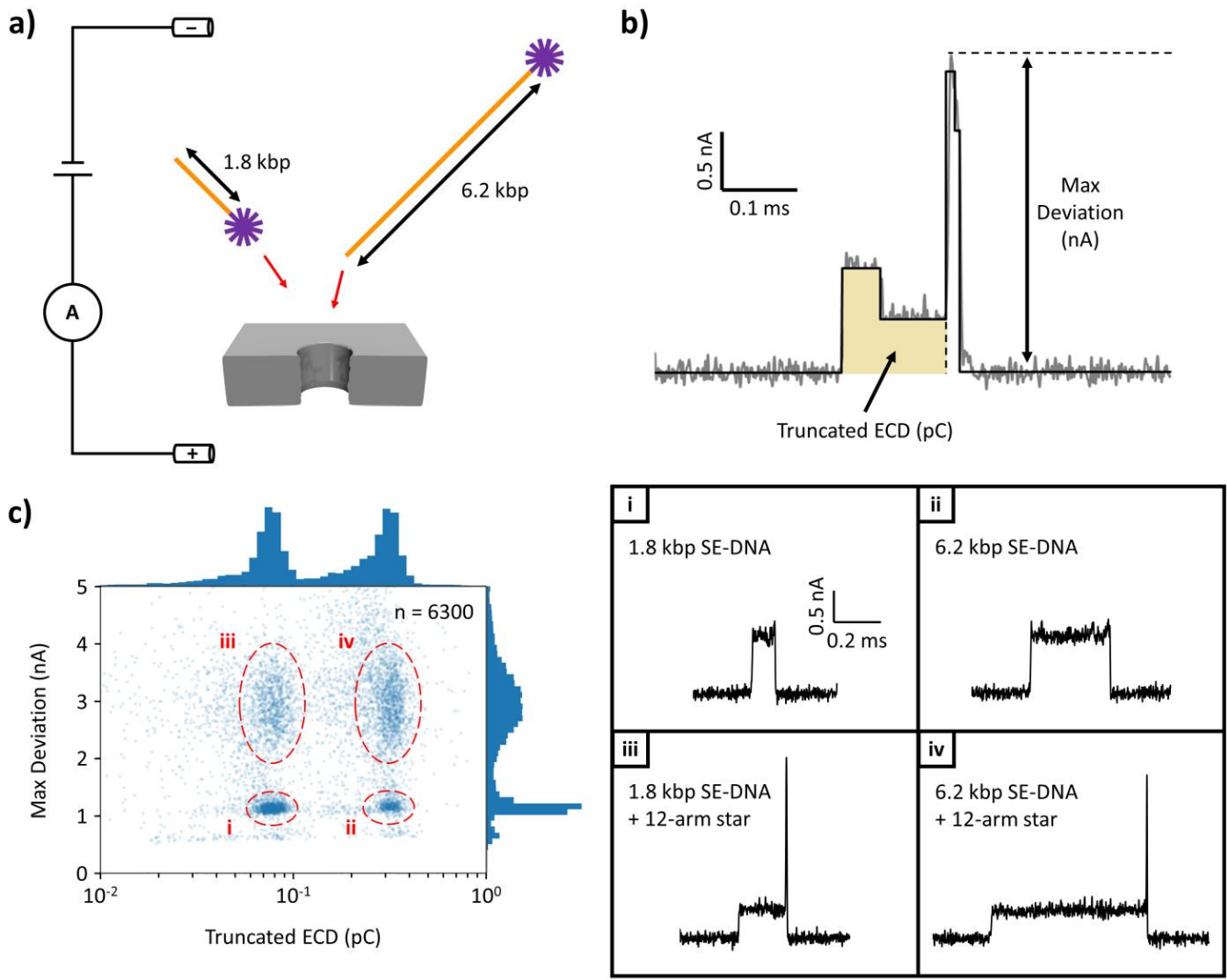

*Figure 2*: *a)* Schematic of the experimental set-up for nanopore detection of a DNA nanostructure attached to multiple SE-DNA lengths. 1.8 kbp and 6.2 kbp SE-DNA molecules were first combined with 12-arm DNA stars via the hybridization of complementary single-stranded extensions from each piece. Individual (negatively-charged) hybrid DNA structures were then electrophoretically driven through a nanopore, generating transient fluctuations ("events") in a background ionic current while they obstruct the pore. *b)* Sample translocation event (current blockage vs. time) illustrating calculations of the event statistics "max deviation" and "truncated ECD" (TrECD). *c)* Scatter plot of max deviation vs. log[TrECD] for events from a single nanopore experiment (~13 nm pore diameter, 3.6 M LiCl pH 8 buffer, 150 mV transmembrane potential). Two main peaks are visible in the histograms (along the margins of the scatter plot axes) of both statistics, resulting in four total event populations corresponding to: i) 1.8 kbp SE-DNA on its own, ii) 6.2 kbp SE-DNA on its own, iii) 1.8 kbp SE-DNA attached to a 12-arm star, iv) 6.2 kbp SE-DNA attached to a 12-arm star. Representative current traces of events from each of the four populations are presented in the panels on the right.

**VALIDATION**

To test the viability of sticky-ended DNA produced by these methods to act as molecular carriers in a nanopore experiment, two product lengths were constructed (1842 bp and 6216 bp) that shared a λ-DNA template, KpnI and SacI restriction enzymes, and a single linker molecule, but that differed in their PCR primers, as outlined in the previous section. The particular linker used here resulted in two complementary ssDNA extensions (5', 12 nt long) being created on either end of each product – see Section S1 in the Supporting Information for the exact oligo sequences (PCR primers and linkers) used.

As proof-of-concept carrier targets, DNA multi-way junctions were utilized in this work. These DNA nanostructures have previously been demonstrated to generate recognizable translocation signals on a nanopore [16,26,27]. Here, 12-way junctions ("12-arm stars") were designed such that each junction arm was 24-bases long, with one of the arms featuring a 12-nt single-stranded extension that is complementary in sequence to a sticky end on the carrier. Incubating a sample of carriers with an excess ($\gtrsim$ 30×) of complementary DNA stars then resulted in the majority of carrier molecules having a target star attached to one end (see Section S6 in the Supporting Information for an example gel shift assay of this attachment step).

A schematic of the nanopore set-up used to detect these carrier-target constructs is illustrated in Figure 2a. Under this scheme, a DNA sample is first diluted in a high-ionic strength buffer (3.6 M LiCl, pH 8) and injected on one side of the pore-containing membrane. A voltage applied across the membrane is then used to transport the negatively-charged molecules though the pore by electrophoresis. As individual molecules approach and translocate though the pore, they block portions of a background flow of ions ($Li^+$ and $Cl^-$ in this case) that are also being electrophoretically driven across the membrane. Molecular translocations can thus be detected with high-bandwidth amplifiers through the transient current fluctuations ("events") they create, signalling the presence of the target molecules in a sample and allowing for their characterization through the details of their current signatures [28,29].

Figure 2b shows a sample current trace of such a translocation event (here resulting from a 12-arm star attached to a 1.8-kbp carrier and passing through a ~13-nm pore), where the amount of ionic current being blocked over time is displayed. Overlaid on top of the raw blockage values (in grey) are a set of fitted step functions ("sublevels", in black) that relate to different substructures of the molecule passing sequentially through the pore [30,31]. In this case, the sequence consists of a baseline (open pore) sublevel, followed by two relatively shallow and long-lived sublevels (corresponding to the extended DNA carrier), then two relatively deep and short-lived sublevels (corresponding to the bulky DNA star), and finally a return to baseline. This suggests that this particular structured molecule was initially captured by its carrier end (specifically in a hairpin configuration given the blockage depth of the first level being ~double that of the second [9,32]) which then threaded through the pore until the bulkier DNA star was reached, resulting in elevated blockage levels above those of the carrier.

Two analysis metrics were employed here to characterize the nanopore signals generated by translocating carrier/DNA star molecules – "max deviation" and "truncated ECD". Max deviation corresponds to the single point of deepest blockage in the signal and is calculated as the difference (in current) between this point and the averaged baseline values (see Figure 2b). As discussed, for hybridized carrier-target molecules, the point of deepest blockage is expected to occur during the translocation of the bulky star attachment and so max deviation indirectly characterizes its size through how much current is blocked [33]. Truncated ECD (for "equivalent charge deficit"), on the other hand, is the integrated area underneath particular fitted sublevels of the event (in between these levels and the baseline) [34]. Here we specifically use the levels attributed to blockages of the carrier within the overall event (see Figure 2b), in order to characterize the length of these carrier molecules through how much total charge they block. Under nanopore conditions where DNA folding is permitted, ECD has been previously been demonstrated to be a more reliable measure of DNA length when compared to overall passage time [18,23,24], since folded sections of the polymer are more compact and so can pass through the pore in shorter times than their fully-extended analogues (in the absence of pore wall interactions).

Figure 2c shows a scatter plot of max deviation vs. truncated ECD (log-scaled) for all carrier translocation events of the same nanopore experiment as the example of Figure 2b (1.8 kbp and 6.2 kbp carriers incubated with 12-arm stars, run though a ~13 nm pore at 150 mV in 3.6 M LiCl pH 8). Excessively short events in duration (< 40 µs) resulting from unsuccessful translocations ("collisions") or the passage of free DNA stars have been removed prior to plotting to reduce clutter (see Section S7 in the Supporting Information). From either the scatter plot itself or from 1D histograms of each metric plotted on the axis margins, at least four main event populations can be identified that are easily associated with the particular configurations of translocating molecules that produced them. The two peaks in truncated ECD, at $77^{+22}_{-17}$ fC and $312^{+99}_{-75}$ fC, correspond to the two different carrier lengths of 1.8 kbp and 6.2 kbp present in the sample. Meanwhile, there is a broad peak (2.9 ± 1.0 nA) at higher values of max deviation and a narrow peak (1.1 ± 0.1 nA) at lower max deviations corresponding to carriers with and without attached stars, respectively. The large spread in max deviation for signals with a star subevent is likely a combined effect of the many possible configurations available to the 12 arms of a translocating star as well as of the variable attenuation of these fast subevents due to the limited bandwidth of the system [26,33]. Together, the event distributions in max deviation and truncated ECD combine to create four main populations in the scatter plot corresponding to: i) free 1.8 kbp carriers, ii) free 6.2 kbp carriers, iii) star-attached 1.8 kbp carriers, and iv) star-attached 6.2 kbp carriers. A representative current trace from each population is presented in the rightmost panels of Figure 2c.

By examining the relative numbers of events in each of the populations (separated using statistic thresholds, see Sections S7 and S8 in the Supporting Information), a few observations can be made. First of all, there are similar numbers of events resulting from 1.8 kbp (3209 events, populations *i* & *iii*) and 6.2 kbp (3061 events, populations *ii* & *iv*) carriers. This reflects the roughly equal concentrations of both carrier types (~0.4 nM each, as estimated by gel electrophoresis) present in the sample mix introduced to the nanopore. Moreover, within a single carrier type, a majority of events have max deviation values consistent with the attachment of a target star, although the exact ratios differed in the two cases – 58% target-attached (*iii*) / 42% free carriers (*i*) for 1.8 kbp and 79% target-attached (*iv*) / 21% free carriers (*ii*) for 6.2 kbp. This discrepancy between carrier lengths could result from the particular (and arbitrary) choice of sticky ends used here, specifically each molecule containing a set of cohesive ends that can anneal to each other to form a closed circle. The shorter the contour length of a carrier molecule, the more often these two ends will encounter each other in the random conformations they adopt before a target star can be attached, leading to an elevated proportion of shorter length carriers that are bound in a circular state [35]. Indeed, looking at the current traces of individual events from the free carrier populations (e.g. see panels *i* and *ii* on the right of Figure 2c), a large fraction contain only a single blockage level at twice the depth of unfolded DNA, consistent with circularized molecules which can only pass through the pore in an even number of dsDNA fragments (see also Section S8 in the Supporting Information for histograms of average blockage for each population). We note however, that whether this circularization effect is viewed as beneficial or detrimental depends on the specific design and goals of a particular nanopore experiment involving these sticky-ended molecules, and that carriers can easily be constructed with only one sticky end or with two ends of non-complementary sequences using the synthesis methods presented here if circularization is to be avoided.

**CONCLUSION**

In summary, a method for the *in vitro* production of sticky-ended DNA was presented for use with nanopore sensing experiments. During the protocol design process, emphasis was placed on the capacity to synthesize products of variable lengths and the ability to carry out individual protocol steps with common molecular biology equipment and widely-available reagents. Our method was successful in producing molecules that generate robust, easy-to-interpret nanopore signals (Figure 2) by virtue of being linear, double-stranded DNA at their cores (whose translocations have been thoroughly studied in the past) and which feature single-stranded ends through which application-specific functionality can be integrated. Moreover, the 12-nt length of the particular set of (cohesive) ends tested here was observed to represent an good balance of creating stable hybridized constructs under standard nanopore sensing conditions (e.g. temperature, buffer composition) while also offering the ability to quickly dissociate these constructs and re-set the system simply by moderate cycling of the temperature (Figure 1b). Lastly, our protocol was designed to be relatively cost- and time-effective when compared to alternative approaches to DNA carrier production. For instance, in contrast to the DNA origami-like approaches of others [1,3,36], which require the purchasing and pipetting of on the order of ~100 oligonucleotide species, the methods presented here make use of a substantially reduced set of reagents, and where only minimal changes to these reagents are needed to make particular product modifications (e.g. replacing a single primer to create a range of product lengths, as in S3 in the Supporting Information). Additionally, after using a viral template for an initial run-though, DNA substrates can thereafter be produced in-house by PCR amplification with our protocol, eliminating the need to continuously purchase commercial stocks of DNA.

In the future, the core methods presented here can also easily be adapted through the incorporation of modified dNTPs or oligos [21,22] during carrier synthesis to meet the needs of future applications that require specific functional groups along their length. The flexibility to integrate further functionality into these molecules post-assembly also exists through the ssDNA overhangs on their ends – these could, for instance, be used as attachment sites for molecular "barcodes" (as defined in the introduction) to more easily distinguish carriers probing for different targets and therefore expand the capability of the system to multiplex. Finally, although molecular carriers have been the focus of this current work, we note that sticky-ended DNA, as a basic construct, may have wider applications to nanopore sensing beyond this. For instance, individual DNA monomers could be chained together by their sticky ends (e.g. circularization, concatemerization [37,38]) to produce nanopore targets with lengths and structures suited for a particular application, starting from more easily-synthesized (shorter, linear) building blocks. Having a greater number of molecular tools available in this way should only assist in opening new avenues for nanopore sensor research into the future.

**MATERIALS AND METHODS**

**DNA Synthesis**

DNA carriers were synthesized as described in the main text (detailed protocols are presented in Section S2 of the Supporting Information). Summarizing, a subsection of a DNA template (λ-DNA, New England Biolabs) was amplified by PCR (LongAmp polymerase, NEB) using specific primer oligonucleotides (Integrated DNA Technologies) to define the amplicon length. The PCR output was then digested by restriction enzymes (KpnI and SacI, NEB) to produce a central fragment with single-stranded overhangs. Finally, DNA linker oligos (IDT) were ligated (T4 ligase, Thermo Scientific) to the ends of this central fragment via its overhangs to create new single-stranded extensions defined by the linker sequences.

For the DNA multiway junctions ("stars") used as targets for the carriers, these were assembled from individual oligonucleotide strands (IDT), as described previously [26,27]. Briefly, a number of strands matching the number of arms in the final structure were mixed in equimolar ratios, heated to 95 °C, and slowly cooled back to room temperature. The sequences of each strand (48 nt long) are designed to hybridize under these conditions to those of neighbouring strands across two adjacent dsDNA arms (24 bp each). One of the assembled arms is also designed to have a single-stranded extension though which it can be annealed to a complementary extension of the carrier. After assembly, the structures were imaged by polyacrylamide gel electrophoresis, their main gel bands excised, and the purified products re-suspended in solution by electroelution (D-Tube Dialyzer Maxi, EMD Millipore).

**Nanopore Sensing**

Nanopores used in this work were fabricated by the controlled breakdown method (CBD), using previously published protocols [39,40]. In short, a silicon nitride membrane (#NBPXA5004Z-HR, Norcada) is immersed in an electrolyte solution (1 M KCl, pH 8) and a relatively large time-varying voltage (~10 V) is applied across it via Ag/AgCl electrodes, using custom instruments and flow cells similar to products from Northern Nanopore Instruments. A computer-controlled amplifier circuit is used to monitor the current through the membrane and to shut off the applied voltage when a sharp increase in this current is detected, signalling the breakdown of the membrane in a single nm-scale hole, as described in Waugh et al. [40].

Prior to a nanopore sensing experiment, a sample of DNA carriers is incubated with an excess of targets (e.g. ~1.5 nM carriers to ~50 nM stars) to anneal the two pieces together (see S6 in the Supporting Information). This annealed mixture is then equilibrated with a high-conductivity sensing solution (3.6 M LiCl, pH 8) and 40 μL injected (at a final carrier concentration of ~0.4 nM) on one side of a nanopore-containing membrane, where an application of moderate transmembrane potentials (~0.1 V) electrophoretically drives the negatively-charged carrier-target molecules through the pore. A high bandwidth current amplifier (VC100, Chimera Instruments), operating at 1 MHz instrument bandwidth (4.17 MHz sampling rate), is used to detect these molecular translocations as fluctuations in the ionic pore current.

Analysis of the current signals ("events") generated by translocating molecules is carried out in the Nanolyzer software package (Northern Nanopore Instruments), which applies a digital low-pass filter to the current trace (400 kHz in this work), locates the events within the timecourse, and fits their signals to step functions that characterize the event substructure (see Figure 2b) [30,31]. Statistics of the analyzed events (e.g. max deviation, average blockage, dwell time, truncated ECD) were calculated either in Nanolyzer itself, or with custom scripts written in Python using the 'pandas' library.

**ACKNOWLEDGMENTS**

The authors would like to acknowledge the support of the Natural Sciences and Engineering Research Council of Canada (NSERC), through funding from grant #CRDPJ 530554-18.

**DECLARATION OF INTERESTS**

V.T.-C. is CSO of Northern Nanopore Instruments Inc., a for-profit company that provides solid-state nanopore tools and software, including the Nanolyzer software used here. Z.R. declares no conflicts of interest.

# SUPPORTING INFORMATION

# Synthesis of Length-Tunable DNA Carriers for Nanopore Sensing

Zachary Roelen, Vincent Tabard-Cossa*

Department of Physics, University of Ottawa, Ottawa, Ontario, Canada

*Corresponding author: tcossa@uottawa.ca

**TABLE OF CONTENTS**

**Section S1: Oligonucleotide sequences**

***Table S1:*** *List of oligonucleotide sequences used to create the sticky-ended DNA products in the main text, using λ-DNA as a PCR template. All oligos were ordered without a 5' terminal phosphate group, especially important in the case of the linker sequences for preventing the permanent formation of circles/concatemers during the ligation step (see Fig. 1a in the main text). Bases highlighted in red and green represent sequences that will eventually constitute (3') overhangs that are compatible with those from digestion with KpnI and SacI restriction enzymes, respectively. In particular, the bolded adenine in the forward primer for the 1.8-kbp product represents a base that differs from the native sequence of λ-DNA and is used to introduce a SacI recognition site in the PCR amplicon. Finally, the underlined sequences in the linker oligos represent regions of complementarity used to hybridize the two strands into an assembled linker molecule.*

| Category | Name | Sequence |
|---|---|---|
| 1.8-kbp Primers | 'FWD-1.8kbp' | CAGTATGG**A**GCTCGGTGGTGTG |
| | 'RVS-1.8kbp' | AGGAACACGGCTCACTTTTACCTT |
| 6.2-kbp Primers | 'FWD-6.2kbp' | GGGACGCTCAGTAATGTGACGATA |
| | 'RVS-6.2kbp' | GGAACTCCGGGTGCTATCAGTTTT |
| Linkers | 'cosL-KpnI' | GGGCGGCGACCTGTAC |
| | 'cosR-SacI' | AGGTCGCCGCCCAGCT |

**Section S2: Detailed SE-DNA synthesis protocols**

A) Polymerase Chain Reaction

1. Combine on ice:
   - 3 μL dNTPs (10 mM, New England Biolabs)
   - 4 μL forward primer (10 μM, Integrated DNA Technologies)
   - 4 μL reverse primer (10 μM, IDT)
   - 1 μL λ-DNA template (20 ng/μL, NEB)
   - 4 μL LongAmp Taq polymerase (2.5 U/μL, NEB)
     20 μL LongAmp reaction buffer (5×, NEB)
   - 64 μL nuclease-free water (Invitrogen)

2. Place in thermocycler:

   | Time | Temperature | Cycles |
   |---|---|---|
   | 30 s | 94 °C | ×1 |
   | 15 s | 94 °C | |
   | 30 s | 60 °C | ×30 |
   | 5 min 20 s | 65 °C | |
   | 10 min | 65 °C | ×1 |

3. Recover DNA product with spin column (e.g. PureLink PCR Purification Kit, Invitrogen) per manufacturer's instructions

B) <u>Double Restriction Digest</u>

1. Combine on ice:
    - 16 µL PCR product (~500 ng/µL, from reaction A)
    - 2 µL KpnI-HF restriction enzyme (20 U/µL, NEB)
    - 2 µL SacI-HF restriction enzyme (20 U/µL, NEB)
    - 10 µL CutSmart reaction buffer (10×, NEB)
    - 70 µL nuclease-free water (Invitrogen)

2. Incubate at 37 °C for 1+ hrs

3. Recover DNA product with spin column (e.g. PureLink PCR Purification Kit, Invitrogen) per manufacturer's instructions

C) <u>Linker Ligation</u>

1. Anneal linker strands by combining 1 µL linker oligo 1 (100 µM, IDT), 1 µL linker oligo 2 (100 µM, IDT), 1 µL T4 ligase reaction buffer (10×, Thermo Scientific), and 7 µL nuclease-free water (Invitrogen) in a tube and then heating for 2 min at 95 °C, 10 min at 50 °C, and cooling to room temperature.

2. Combine on ice:
    - 55 µL digested DNA (~70 ng/µL, from reaction B)
    - 10 µL annealed linkers (10 µM, from step 1)
    - 4 µL T4 ligase (5 U/µL, Thermo Scientific)
    - 19 µL T4 ligase reaction buffer (10×, Thermo Scientific)
    - 112 µL nuclease-free water (Invitrogen)

3. Incubate overnight at 16 °C

4. Recover DNA product with spin column (e.g. PureLink PCR Purification Kit, Invitrogen) per manufacturer's instructions

## Section S3: Sample distribution of product lengths

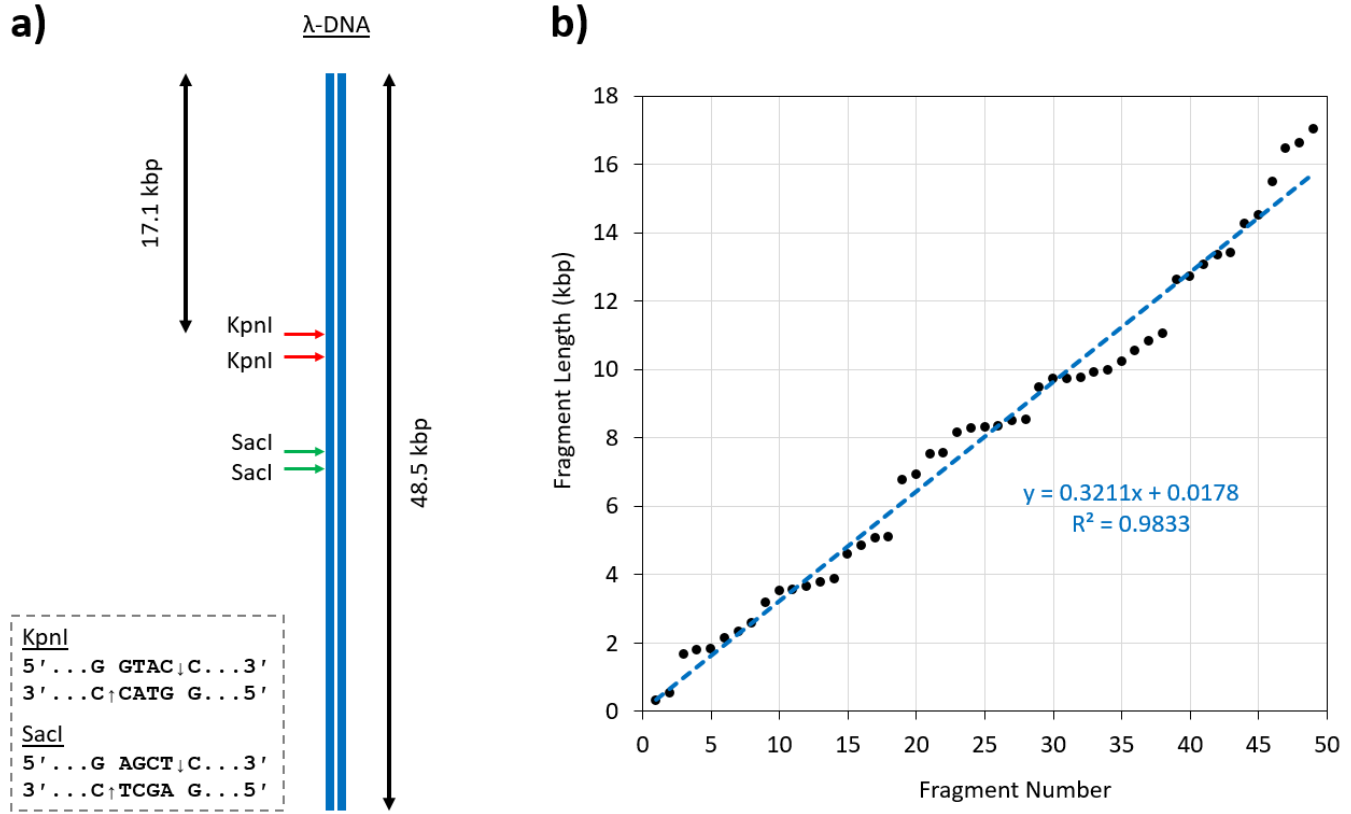

***Figure S3:*** *a) Restriction map of λ-DNA for KpnI (red arrows) and SacI (green arrows) restriction enzymes. Each enzyme has only two (relatively closely-spaced) recognition sites in the entire 48.5 kbp template – see inset for sequences. Labelled here in particular is an initial stretch of 17.1 kbp that is devoid of recognition sites for either enzyme. b) Distribution of possible fragment sizes within this 17.1 kbp region by digest at: 1) the KpnI site bounding the region and 2) new SacI sites introduced by modification of the native λ sequence by a single base, using a mismatched PCR primer (e.g. 'FWD-1.8kbp' in Table S1). A full range of relatively equally-spaced (median gap size of 227 bp) fragments is produced from 0 – 17 kbp, where each fragment will feature two different overhang sequences on their ends (from SacI and KpnI – both 3' polarity, 4-nt) for selective targeting by a linker molecule in the ligation step (see Fig. 1a in the main text). Note that the density of fragment size coverage can be increased (essentially arbitrarily) beyond that of this simple example by loosening the degree of complementarity between the primer and the template (especially by simply adding the SacI recognition sequence to the upstream / 5' end of the primer), or by simultaneously adjusting the location of the KpnI site in a similar manner.*

## Section S4: Agarose gels of short fragment removal

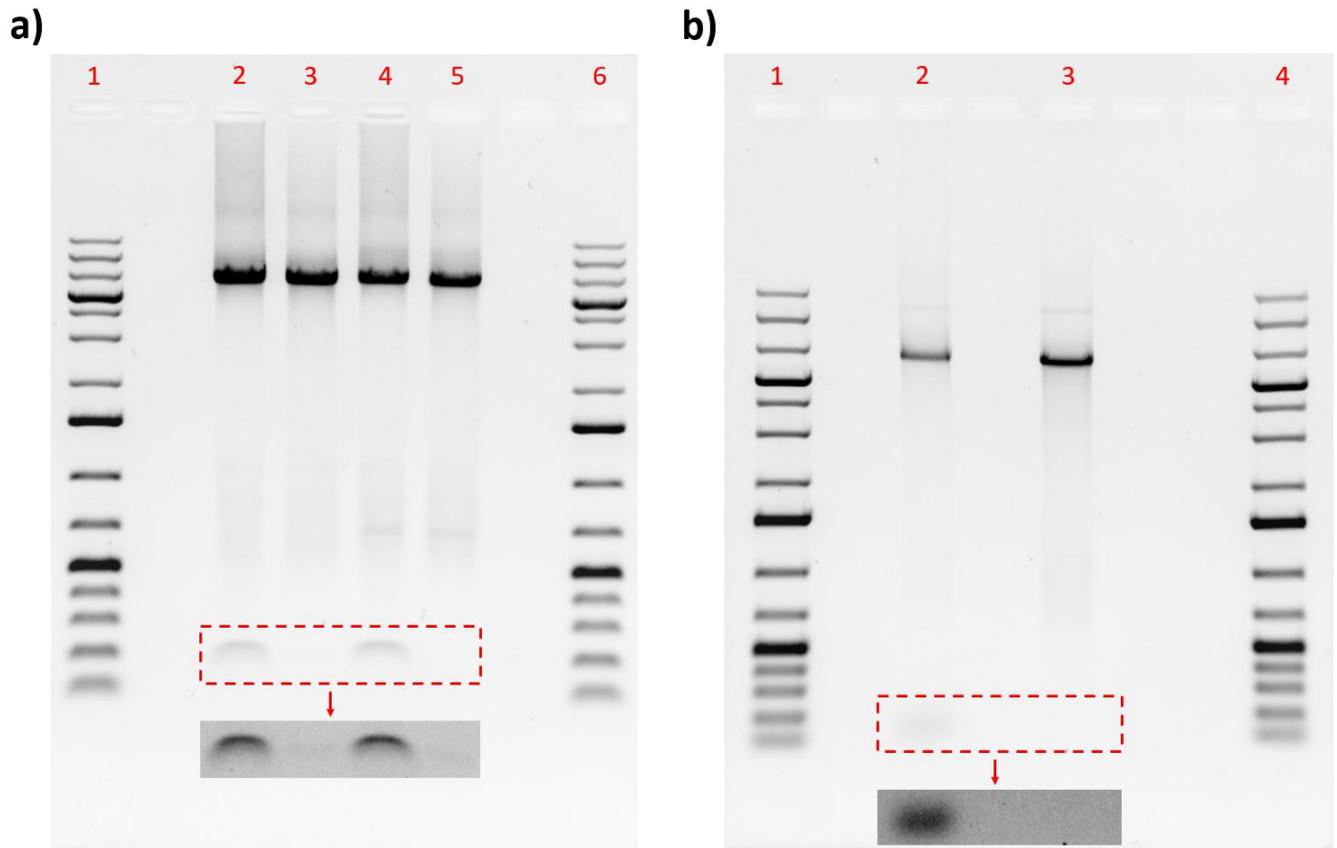

***Figure S4:*** *Removing short spurious fragments during SE-DNA production by spin column purification. a) Agarose gel (0.7%, 0.5× TBE, pre-stained with Gel Red, 70 V) of KpnI-SacI double digest of 6.2 kbp PCR amplicon. Lanes 1 & 6: GeneRuler 1 kb Plus DNA Ladder (Thermo Scientific). Lanes 2 & 4: raw KpnI-SacI digest mix with 6.2 kbp amplicon from two alternate primer sets. Lanes 3 & 5: spin column purification (PureLink Quick PCR Purification Kit, Invitrogen) of Lane 2 and Lane 4 products, respectively. An enhanced-contrast view of the dashed region is presented as an inset, where the sticky-ended "caps" (70 and 144 bp or 49 and 144 bp) of the central digest fragment (see Fig 1a in the main text) are observed to have been largely removed during the purification step. b) Agarose gel (0.6%, 0.5× TAE, pre-stained with Gel Red, 70 V) of T4 ligation of the 6.2 kbp double-digest product from (a) and a "linker" molecule (sequence in Table S1). Lanes 1 & 4: GeneRuler 1 kb Plus DNA Ladder (Thermo Scientific). Lane 2: raw T4 ligation mix. Lane 3: spin column purification (PureLink Quick PCR Purification Kit, Invitrogen) of Lane 2 products. An enhanced-contrast view of the dashed region again shows the removal of short spurious DNA fragments, this time of excess linkers (16 bp). By keeping the DNA by-products produced/present at each synthesis stage short in length (and thus easily removable) though careful sequence design, they will not be available to interfere in subsequent synthesis stages by re-annealing with the main product.*

## Section S5: Nanopore sensing of agarose-extracted DNA

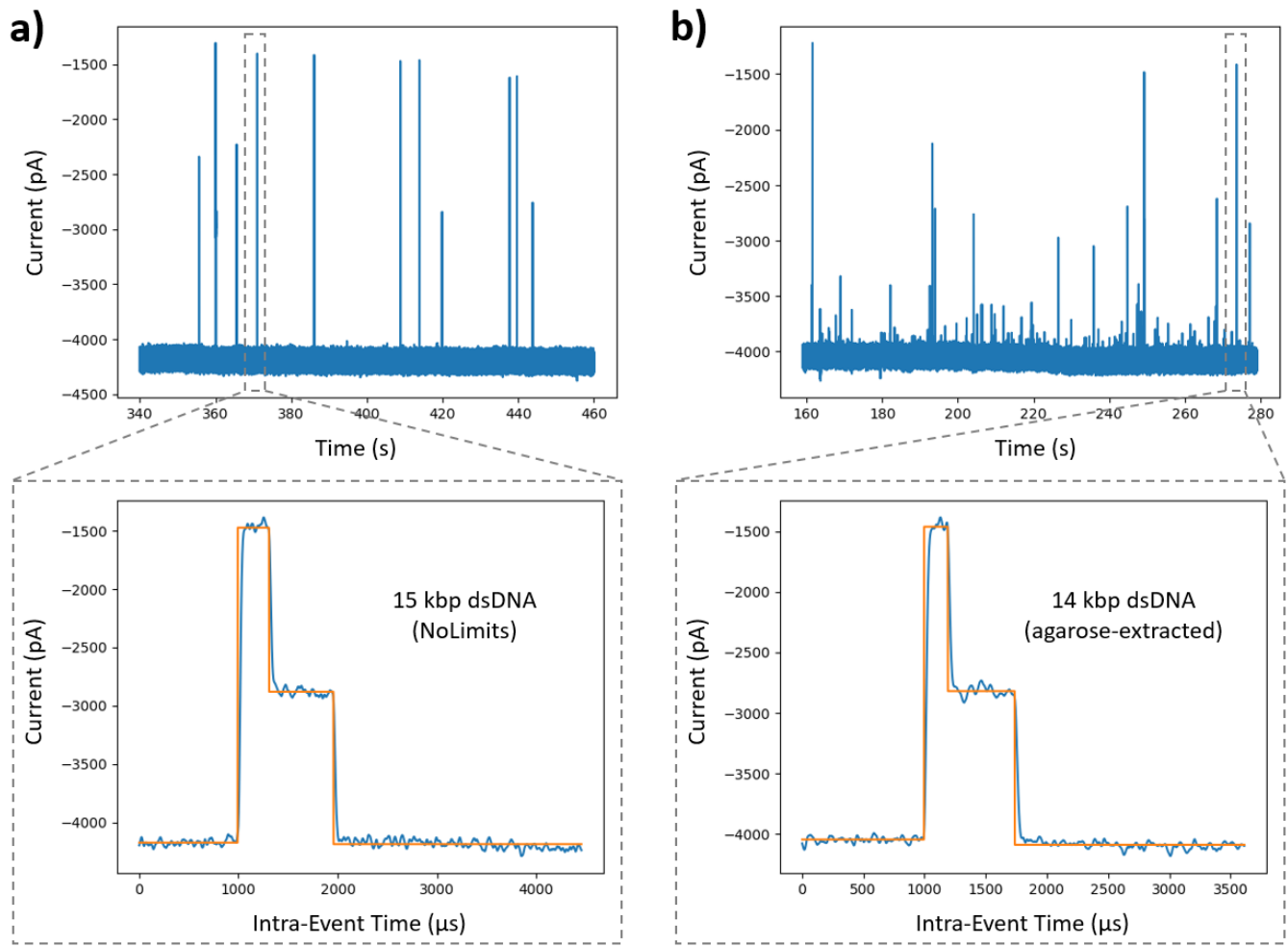

***Figure S5:*** *Comparing nanopore signals (~4.9-nm pore diameter, 200 mV transmembrane potential, 3.6 M LiCl pH 8 buffer) produced by commercial and agarose-extracted DNA samples. a) Representative (2 minute) current trace of a 15 kbp commercial dsDNA sample (NoLimits, Thermo Scientific) passing through the pore. b) Representative (2 minute) current trace of a 14 kbp dsDNA sample (BstEII digest fragment of λ-DNA) extracted from an agarose gel (GeneJET Gel Extraction Kit, Thermo Scientific) passing through the same pore as in (a). While both samples produced recognizable dsDNA signals (example traces expanded as insets), agarose-extracted DNA reliably led to the appearance of additional, lower amplitude signals (see S5b) as if from contaminants, as well as increased pore clogging (both reversible and irreversible with voltage inversion or solution flushing). This behaviour persisted across all of the numerous extraction methods tested including gel dissolution (e.g. QIAquick, QIAGEN), centrifugal separation (e.g. Freeze 'N Squeeze, Bio-Rad), enzymatic digestion (e.g. agarase, NEB), and passive diffusion into supernatant out of submerged gel pieces. In contrast, DNA synthesis protocols that avoided an agarose-extraction step (but with otherwise similar sample manipulations), produced molecules that resulted in longer-lasting, cleaner-to-analyze nanopore experiments (such as presented in Figure 2 of the main text). Alternative approaches to isolating specific DNA fragments were thus implemented in this work (see Section S4).*

## Section S6: Agarose gel of carrier/DNA star hybridization

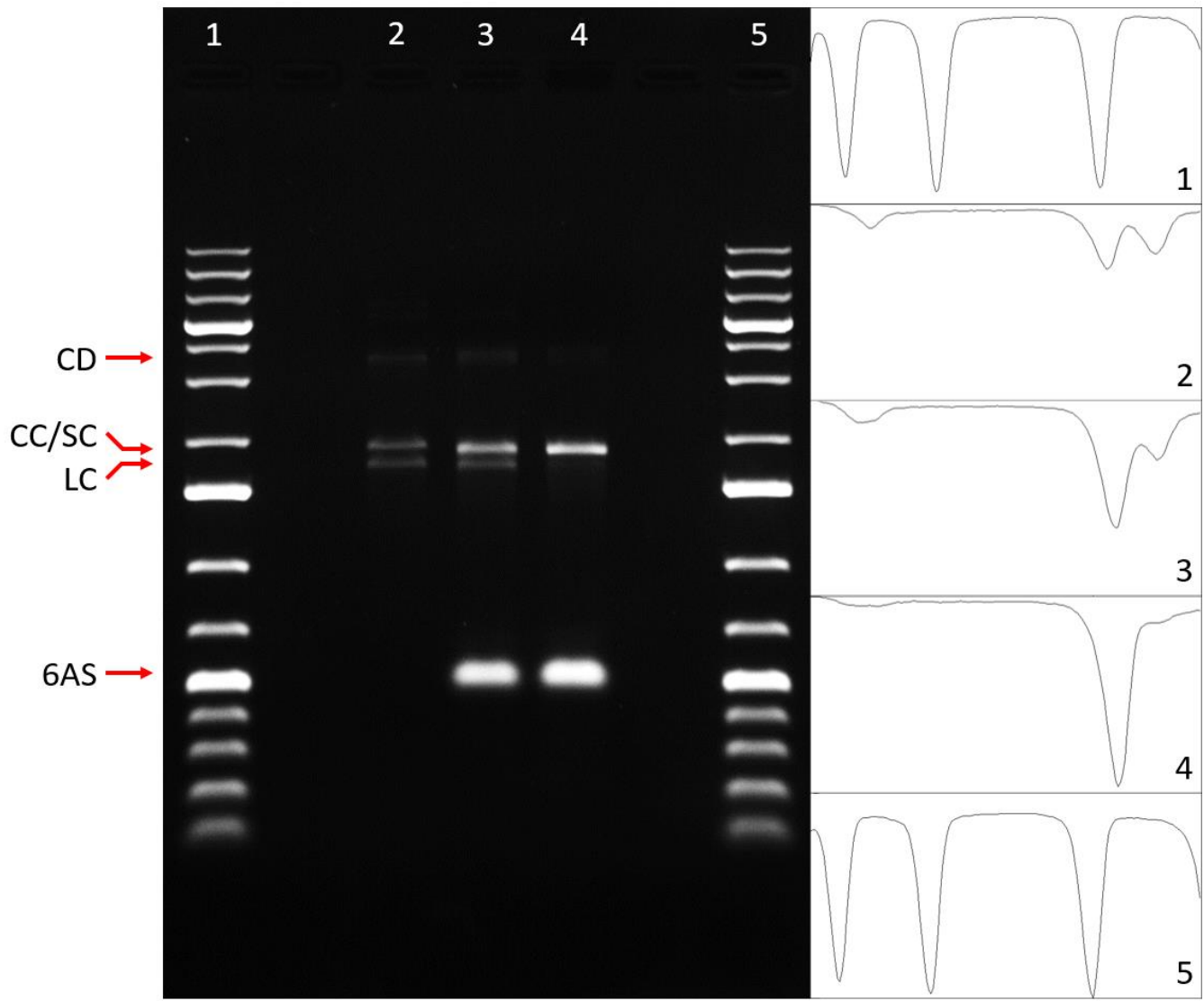

***Figure S6:*** *Agarose gel (0.7%, 0.5× TBE, pre-stained with Gel Red, 70 V) showing the hybridization of 1.8-kbp carriers to 6-arm stars via complementary single-stranded extensions on both pieces. Lanes 1 & 5: GeneRuler 1 kb Plus DNA Ladder (Thermo Scientific). Lane 2: 1.8-kbp SE-DNA incubated for 1 hr at 37 °C (10 mM MgCl$_2$ + 10 mM Tris, pH 8). Lane 3: 1.8-kbp SE-DNA incubated with 6-arm stars (at ~100× excess) for 1 hr at 37 °C. Lane 4: 1.8-kbp SE-DNA heated 10 min at 70 °C, flash chilled, then incubated with 6-arm stars (at ~100× excess) for 1 hr at 37 °C. Four main bands are visible across the product lanes and are labelled "6AS" (6-arm stars), "LC" (linear carriers), "CC" (circular carriers), "SC" (star-annealed carriers), and "CD" (carrier dimers) after their presumptive molecular identities. The panels on the right show the integrated band intensities at the positions of the top three (carrier-associated) bands for each lane. As the same mass of carrier was loaded into all product lanes, the increase in intensity of the CC/SC band in lane 3 relative to lane 2 (lacking DNA stars) is attributed to the attachment of stars to a portion of the carriers. The further intensity increase of this band in lane 4 (combined with the simultaneous reduction in the LC band) when the star incubation was preceded by heating the carriers above the melting temperature of their ends implies that some fraction of the carriers were inaccessible to the stars (e.g. circularized or attached to stray linker oligos) prior to this melting step. Such a step was therefore incorporated into subsequent annealing reactions when producing analytes for nanopore sensing.*

## Section S7: Filtering out short-lived nanopore events

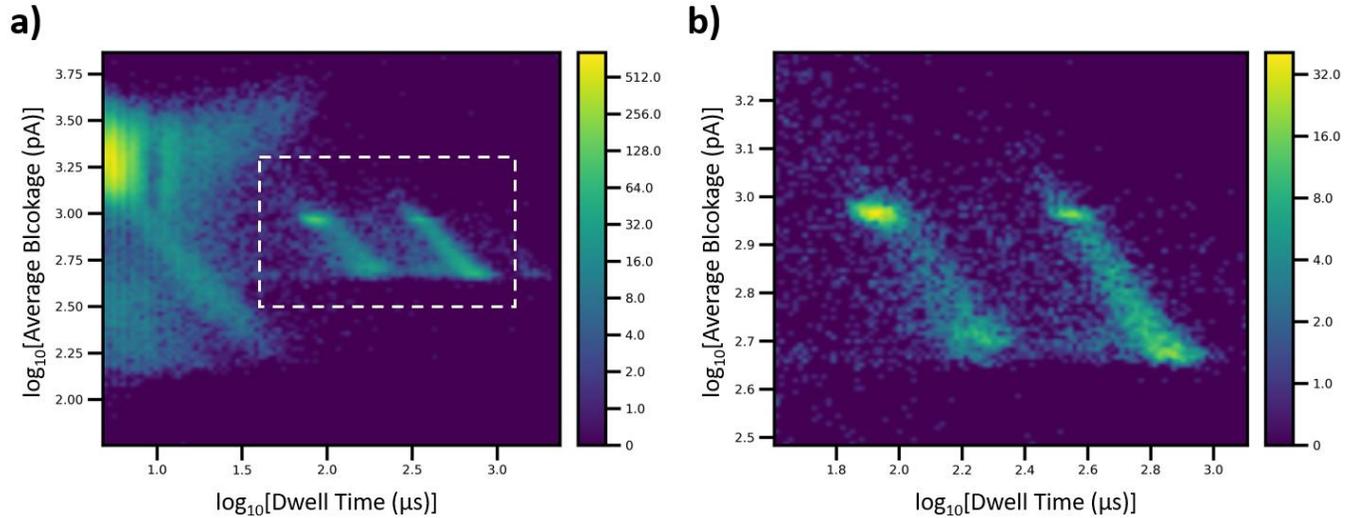

***Figure S7:*** *a) 2D histogram of average blockage vs. dwell time (log-log scaling) for all nanopore events from the experiment represented in Figure 2 of the main text (1.8- and 6.2-kbp DNA carriers, 12-arm stars run on ~13 nm pore in 3.6 M LiCl pH 8 buffer, 150 mV transmembrane potential). Two main populations of long-lived events are visible on the right side of the plot, corresponding to translocations of the two carrier sizes (attached to a DNA star or otherwise). This contrasts with the diffuse mass of short-lived events on the left side of the plot, widely-spread in average blockage, which likely correspond to failed pore entries ("collisions") as well as to translocations of free 12-arm stars (added in large excess to the carriers), especially those events with the deepest average blockages. A set of event selection filters, represented by the dashed box in S7a, was applied to the data to only consider events from the carriers during analysis: dwell time $\in$ [40 µs, 1300 µs] & average blockage $\in$ [300 pA, 2000 pA]. b) 2D histogram of average blockage vs. dwell time (log-log scaling) for this filtered set of events, showing the isolation of the two main carrier populations. This reduced subset was then used for the analysis of carrier translocations shown in Figure 2 of the main text.*

## Section S8: Average blockages of Figure 2 populations

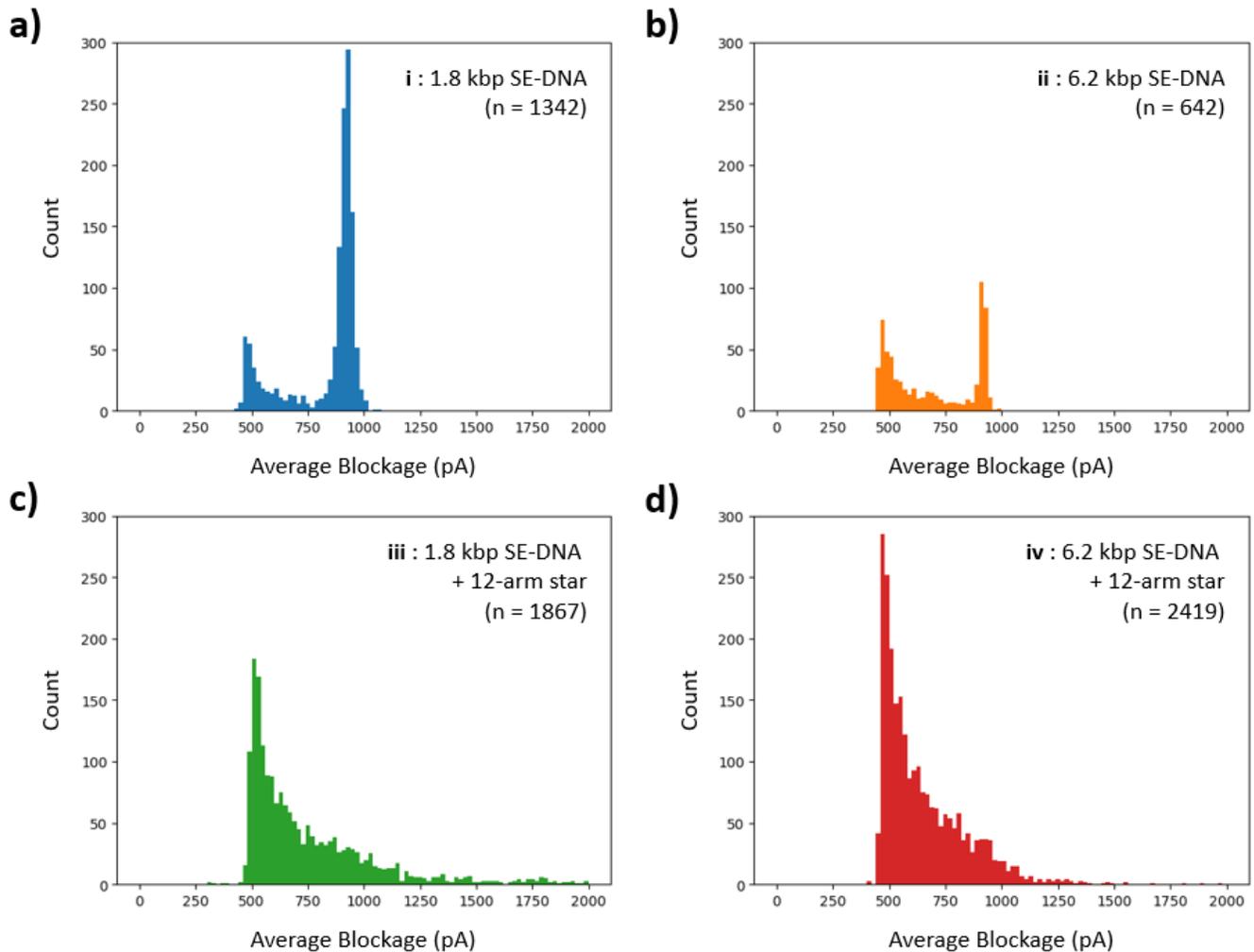

*Figure S8:* 1D histograms of average blockage for the events of Figure 2 in the main text (1.8- and 6.2-kbp DNA carriers, 12-arm stars run on ~13 nm pore in 3.6 M LiCl pH 8 buffer, 150 mV transmembrane potential), separated by the population types (i, ii, iii, iv) of Figure 2. The events were sorted by population using selection filters as outlined in Fig. S7, this time on max deviation and truncated ECD. a) Average blockage distribution of population i (1.8 kbp SE-DNA). b) Average blockage distribution of population ii (6.2 kbp SE-DNA). c) Average blockage distribution of population iii (1.8 kbp SE-DNA annealed to 12-arm star). d) Average blockage distribution of population iv (6.2 kbp SE-DNA annealed to 12-arm star). The star-attached populations (iii, iv) show a slow decay of average blockages from the single dsDNA level (~450 pA, fully unfolded events) to the double dsDNA level (~900 pA, fully folded events) as these molecules get captured closer and closer to the middle of their contour lengths. The free carrier populations (i, ii), on the other hand, show sharp peaks at the double dsDNA level, implying that an elevated proportion translocate in a fully-folded conformation. This is consistent with many of the free carriers having their ends annealed together to form a closed circle – this forces the molecules to pass through the pore two dsDNA fragments at a time and blocks their annealing to complementary target stars.